\documentclass[]{aa}
\usepackage[colorlinks=true, citecolor = blue, urlcolor = blue, linkcolor = blue]{hyperref}
\usepackage{graphicx}
\usepackage{epstopdf}

\begin{document}

\title{Formation of the thermal infrared continuum in solar flares}

\author{
        Paulo J. A. Sim\~oes\inst{\ref{inst1}}
    \and 
        Graham S. Kerr\inst{\ref{inst1},\ref{inst5}}
    \and
        Lyndsay Fletcher\inst{\ref{inst1}}
    \and
        Hugh S. Hudson\inst{\ref{inst1},\ref{inst2}}
    \and
        C. Guillermo Gim\'enez de Castro\inst{\ref{inst3}}
    \and    
        Matt Penn\inst{\ref{inst4}}
}

\institute{
SUPA School of Physics and Astronomy, University of Glasgow, Glasgow, G12 8QQ, UK \email{Paulo.Simoes@glasgow.ac.uk} \label{inst1}
\and
Space Sciences Laboratory, University of California, Berkeley, CA, USA \label{inst2}
\and 
Centro de R\'adio Astronomia e Astrof\'isica Mackenzie, Rua da Consola\c{c}\~ao 896, 01302-907, S\~ao Paulo, Brazil\label{inst3}
\and
National Solar Observatory, Tucson, AZ 85719, USA \label{inst4}
\and
Current address: NASA Goddard Space Flight Center, Heliophysics Sciences Division, Code 671, 8800 Greenbelt Rd., Greenbelt, MD 20771, U.S.A. \label{inst5}
}

\titlerunning{Formation of IR continuum in flares}

\authorrunning{Sim\~oes et al.}

\date{Received / Accepted}


\abstract{}
{Observations of the Sun with the Atacama Large Millimeter Array have now started, and the thermal infrared will regularly be accessible from the NSF's Daniel K. Inouye Solar Telescope. Motivated by the prospect of these new data, and by recent flare observations in the mid infrared, we set out here to model and understand the source of the infrared continuum in flares, and to explore its diagnostic capability for the physical conditions in the flare atmosphere.}
{We use the 1D radiation hydrodynamics code RADYN to calculate mid-infrared continuum emission from model atmospheres undergoing sudden deposition of energy by non-thermal electrons.}
{We identify and characterise the main continuum thermal emission processes relevant to flare intensity {enhancement} in the mid- to far-infrared (2--200~$\mu$m) spectral range as free-free emission on neutrals and ions. We find that the infrared intensity evolution tracks the energy input to within a second, albeit with a lingering {intensity enhancement}, and provides a very direct indication of the evolution of the atmospheric ionization. The prediction of highly impulsive emission means that, on these timescales, the atmospheric hydrodynamics need not be considered in analysing the mid-IR signatures. 
}{}

\bibpunct{(}{)}{;}{a}{}{,} 

\maketitle

\section{Introduction} \label{sec:intro}

Solar flares are the product of sudden energy release in compact regions in the solar atmosphere. The energy transferred from the magnetic field in active regions to the local plasma causes heating and particle acceleration, leading to the emission of electromagnetic radiation. The solar flare phenomenon was first discovered independently by \citet{Carrington:1859} and \citet{Hodgson:1859}, in visible light, and since then, the broadband nature of the electromagnetic emission of flares became clear, following the first detection of flare emission in several spectral ranges, radio \citep{Hey:1983}, hard \citep{PetersonWinckler:1959} and soft X-rays \citep{White:1964}, $\gamma$-ray lines \citep{ChuppForrestHigbie:1973,TalonVedrenneMelioranskii:1975}. 
 Sporadic observations in the infrared (IR) and {sub-millimetric (sub-mm) wavelengths have been performed over the years, but not aimed at flares studies \citep[see ][for a summary of these observations]{LoukitchevaSolankiCarlsson:2004}.} 
Only recently, however, observations in the mid-infrared range become feasible for flare studies \citep{KaufmannWhiteFreeland:2013,PennKruckerHudson:2016}. As we will demonstrate in this paper, flare IR continuum observations are of particular value as they give us a very direct diagnostic of atmospheric electron content, with a temporal and spatial resolution adequate to discriminate between different flare energy transport models.
 
The first observations of flares in the IR continuum revealed flare ribbons at 1.56 $\mu$m in two strong X-class events, SOL2003-10-29 (X10) and SOL2003-11-02 (X8) \citep[][]{XuCaoLiu:2004,XuCaoLiu:2006}. 
Since then, observations at longer wavelengths have been  reported:  SOL2012-03-13 (M8) at 10~$\mu$m (30 THz) \citep{KaufmannWhiteFreeland:2013}, SOL2014-03-29 (X1) at 1.08~ $\mu$m, near the He~{\sc i} 10830 \AA~line \citep{KleintHeinzelJudge:2016}. 
The first spectral mid-IR flare observation, in two broad bands at 5.2 and 8.2 $\mu$m, was reported by \citet{PennKruckerHudson:2016}, {for the event SOL2014-09-24 (C7.0)}. 
These observations revealed  the mid-IR sources to be compact and consistent with double-footpoint or ribbon configurations, well-associated with hard X-ray (HXR) and white light (WL) sources.
The observed mid-IR emission has a strongly impulsive character.

The first tentative description of IR flare emission was proposed by \citet{OhkiHudson:1975} as a part of early unsuccessful searches for flare signatures \citep[cf.][]{Hudson:1975}.  
They suggested two main possible emission mechanisms: optically thin thermal free-free radiation from the chromosphere or blackbody radiation (dominated by H$^-$ opacity) from the heated photosphere, following the ideas proposed to explain the formation of the white-light continuum in flares. 

More recently, \citet{HeinzelAvrett:2012} performed calculations of thermal IR flare emission using existing semi-empirical model atmospheres of flares \citep{MachadoAvrettVernazza:1980,MauasMachadoAvrett:1990}. 
Their results strongly indicated the dominance of ion free-free emission from the chromosphere.\footnote{We refer to free-free emission mechanism on ions, predominantly H~{\sc ii}, as ``ion free-free'', and on neutrals as ``neutral free-free'' continua.
The latter is often also called ``H$^-$ free-free emission'' for hydrogen.
Generally free-bound contributions are relatively small in the mid-IR.} 
A similar method was employed by \citet{TrottetRaulinMacKinnon:2015} to investigate the origin of the 30 THz (10 $\mu$m) emission in the M-class flare SOL2012-03-13 \citep{KaufmannWhiteFreeland:2013}, and led to similar conclusions. \citet{PennKruckerHudson:2016} have shown that the observed IR spectrum {during SOL2014-09-24} is consistent with free-free emission from a chromospheric region with an upper limit to the electron density of $\approx 4 \times 10^{13}$ cm$^{-3}$. \citet{KasparovaHeinzelKarlicky:2009} used non-LTE radiative hydrodynamic models to investigate the effects of non-thermal excitation and ionisation of hydrogen in the formation of {sub-mm} emission. \citet{ChengDingCarlsson:2010} used radiative hydrodynamic simulations performed with RADYN \citep{CarlssonStein:1995,CarlssonStein:1997} to investigate the formation of the continuum at 1.56~$\mu$m \citep{XuCaoLiu:2006}. \citet{KleintHeinzelJudge:2016}, using observations in the near IR, optical and near ultraviolet (NUV), have concluded that the NUV contiuum excess intensity is more consistent with H free-bound continuum, while the optical and near-infrared contiuum excess intensity could be explained by a blackbody spectrum.

In this work, we employ 1D radiative hydrodynamic simulations using RADYN to investigate the formation of the IR continuum in the dynamically evolving flare atmosphere.
We study very short episodes of energy injection, to be able to isolate the most important timescales namely ionization and recombination.
This important limit has previously been studied by \citet{KasparovaHeinzelKarlicky:2009}, who showed the possibility of sub-second emission signatures in the sub-mm ranges.
We find that the continuum IR signature, particularly at longer wavelengths, tracks energy injection with sub-second fidelity, and that the only parameter of importance is forming the radiation is atmospheric ionization. This means that details of the hydrodynamic development of the atmosphere are irrelevant for interpreting this measurement on short timescales, consistent with the simplifying ``constant density'' limit of  \citet{ShmelevaSyrovatskii:1973}, in which the atmospheric structure does not change while the radiation is being formed.
We present a detailed analysis of the formation height of the IR continuum, main sources of opacity, synthetic spectra and light curves.  We also compare our results with IR data from \citet{PennKruckerHudson:2016}.

\section{Calculations of the infrared continuum in solar flares} \label{sec:method}

The formation of the infrared continuum is significantly simpler than the formation of recombination continua or spectral lines because the continuum source function is given by the local Planck function $S_\nu=B_\nu(T)$, with $T$ being the kinetic (or electron) temperature. Nonetheless, the chromospheric opacities depend on electron, proton and neutral hydrogen densities that must be treated under fully non-LTE conditions. In this work, we used RADYN \citep{AllredHawleyAbbett:2005,AllredKowalskiCarlsson:2015} simulations to obtain the dynamic flaring chromosphere, subjected to collisional heating by a non-thermal distribution of electrons, as described in Sect. \ref{sect:models} below.

For the calculations of the IR emission we follow a similar approach to previous works \citep[e.g.][]{HeinzelAvrett:2012} where a 1D model atmosphere is used to calculate the relevant opacities 
and the radiative transfer for a range of IR wavelengths. In this work we calculate the radiation emerging from the top of the 1D atmosphere, i.e. with a heliocentric angle cosine $\mu=1$. Center-to-limb effects are likely to be important; these cannot be assessed with 1D models, as one needs to account for the intrinsic anisotropy of the flare emission, and also its propagation through the quiescent atmosphere surrounding the flare site.

In the chromosphere, the main source of opacity for the infrared continuum is ion free-free continuum, while in the temperature minimum region and below the opacity is dominated by neutral free-free opacity.

The ion free-free opacity $\kappa_\nu (\mathrm{H})$ (in $\mathrm{cm}^{-1}$) is calculated by \citep{RybickiLightman:1986}
\begin{equation}
\kappa_\nu(\mathrm{H}) = 3.7 \times 10^8 T^{-1/2} n_e n_p \nu^{-3} g_\mathrm{ff}
\label{eq:kff}
\end{equation}
where $n_e$ and $n_i$ are the electron and proton densities respectively, and $T$ is the kinetic (or electron) temperature. {Note that $n_e$ includes contributions from hydrogen, {helium and metals}, as calculated by the employed model, see Section \ref{sect:models}.}
{Accurate values for the Gaunt factor $g_\mathrm{ff}$ were obtained by interpolating the table of numerically calculated $g_\mathrm{ff}$ provided by \citet{van-HoofWilliamsVolk:2014}.}

Lower in the atmosphere, around the temperature minimum region and below, H$^-$ free-free opacity dominates. {We use the expression for the H$^-$ free-free opacity given by \citet{Kurucz:1970}.} 

\begin{equation}
\kappa_\nu (\mathrm{H}^-) = \frac{n_en_\mathrm{H}}{\nu}(A_1 + (A_2 - A_3/T)/\nu),
\label{eq:khm}
\end{equation}
where $n_\mathrm{H}$ is the neutral hydrogen density, and the numerical coefficients are $A_1=1.3727 \times 10^{-25}$, $A_2=4.3748 \times 10^{-10}$, and $A_3=2.5993 \times 10^{-7}$ \citep{Kurucz:1970}. {A similar expression given by \citet{John:1988} is in good agreement with Eq. \ref{eq:khm}, within 1--3\%.}

Using Eq. \ref{eq:kff} and \ref{eq:khm}, the total opacity is then
\begin{equation}
\kappa_\nu=[\kappa_\nu(\mathrm{H})+\kappa_\nu(\mathrm{H^-})](1-e^{-h\nu/k_bT})
\end{equation}
with the term $(1-e^{-h\nu/k_bT})$ being the correction for stimulated emission, where $h$ and $k_b$ are the Planck and Boltzmann constants.

Here we define the {\em contribution function} $\mathrm{CF}$ (given in erg s$^{-1}$ cm$^{-3}$ Hz$^{-1}$ sr$^{-1}$ throughout this paper), following e.g. \citet{Carlsson:1998}:
\begin{equation}
\mathrm{CF}(h)=j_\nu e^{-\tau_\nu},
\end{equation}
where $j_\nu=\kappa_\nu B_\nu(T)$ is the emission coefficient, and $B_\nu(T)$ is the Planck function. {We define CF as a function of height $h$ where the height is related to the optical depth scale via $\mathrm{d}\tau_\nu=-\kappa_\nu \mathrm{d}h$, the optical depth.} The contribution function $\mathrm{CF}$ indicates the formation height of emission. 

Finally, the specific intensity is calculated by integrating the $\mathrm{CF}$ along the line of sight:
\begin{equation}
I_\nu=\int \mathrm{CF}(h) \ \mathrm{d}h,
\end{equation}
which we use to calculate the brightness temperature $T_b$ using the Rayleight-Jeans law:
\begin{equation}
T_b(\nu) = \frac{c^2}{2k_b\nu^2}I_\nu.
\label{eq:tb}
\end{equation}

\section{Radiative hydrodynamic flare modelling} \label{sect:models}

RADYN is a well-established numerical tool for modelling solar flares. \citet{AbbettHawley:1999} and \citet{AllredHawleyAbbett:2005} modified the code of \citet{CarlssonStein:1995,CarlssonStein:1997} to simulate flares, and since then RADYN has been used by several authors to better investigate the hydrodynamic and radiative response of the solar atmosphere to the injection of flare energy \citep[e.g.][]{KuridzeMathioudakisSimoes:2015,KennedyMilliganAllred:2015,Rubio-da-CostaKleintPetrosian:2016,KerrFletcherRussell:2016}

RADYN solves the coupled, non-local equations of hydrodynamics, radiation transport and atomic level populations (including time-dependent effects) on a 1D adaptive grid. Backwarming by soft X-ray, extreme ultraviolet and ultraviolet (XEUV) radiation is included self-consistently in the code as an additional heating term. For a recent description of RADYN, along with a list of atomic transitions solved, the reader is encouraged to consult \citet{AllredKowalskiCarlsson:2015}.
In part because of the restrictions of 1D, the results presented here are numerical experiments and should not be interpreted as direct modelling of specific flares.  Instead, in this work we have tried to understand the ways in which the IR emission shows the response of the chromospheric plasma activated by flare energy input. 
\begin{figure*}
\resizebox{\hsize}{!}{\includegraphics[angle=0]{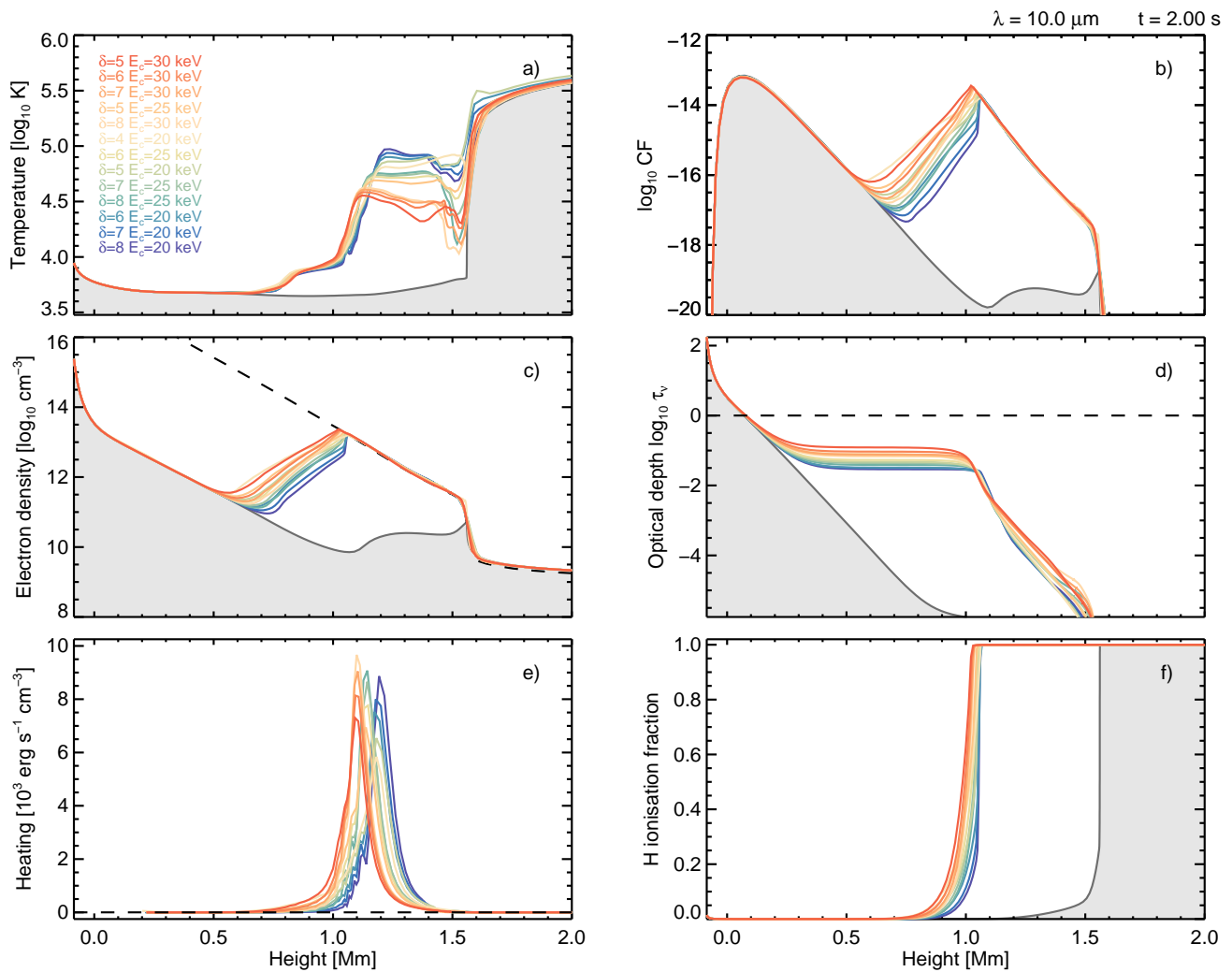}}
\caption{The RADYN quiescent atmosphere (gray) and an overview of all of the model results for 10~$\mu$m continuum, evaluated at peak energy input time (2~s).
Top row, temperature and contribution function (CF).
Middle row, electron density and optical depth.
The dashed line shows all models to lie below $\tau = 1$ at this wavelength.
Bottom row, the heating rate and the resulting ionization fraction for hydrogen.
Model colour coding corresponds roughly to the number of electrons at 50~keV, with red being highest.
}
\label{fig:cf10um}
\end{figure*}
In our use of the RADYN code we explicitly assume electron energy transport, as in the ``thick-target'' model \citep{Brown:1971,Hudson:1972}, in which an electron beam formed in some coronal region precipitates into the pre-flare atmosphere.
We further explore this model only in the restrictive case of vertical, unidirectional beam propagation (but including RADYN's Fokker-Planck treatment of electron propagation), and ignore other forms of energy transport such as the Poynting flux \citep{EmslieSturrock:1982,FletcherHudson:2008}.

The electron beam is described as a power law in energy with spectral index $\delta$ and low-energy cutoff $E_c$. We have run a set of RADYN simulations with spectral index $\delta = 5$, 6, 7, and 8, for each value of the low energy cutoff $E_c= 20$, 25, and 30~keV. 
Here $f(E) = dN/dE \propto H(E_c) E^{-\delta}$, where $H(E)$ is the Heaviside function.
In all 13 models the energy input followed a Gaussian time profile with a {FWHM of 0.553~s} and with a maximum of $F=\int Ef(E) dE=10^{11}\ \mathrm{erg} \ \mathrm{s}^{-1} \ \mathrm{cm}^{-2}$ at $t=2.0$ s. 
{Hereafter, we define our energy input function as F11t, following the typical notation in previous works e.g. F11 for $F=10^{11}\ \mathrm{erg} \ \mathrm{s}^{-1} \ \mathrm{cm}^{-2}$ in \cite{AllredHawleyAbbett:2005}.}
Note that this beam power per unit area roughly matches the solar luminosity per unit area.
In Figure~\ref{fig:cf10um} we summarize the properties of the atmosphere at the time of peak energy deposition.
Different electron beams will heat the atmosphere in slightly different ways \citep[see detailed discussion in][]{AllredKowalskiCarlsson:2015}. 
The low-energy cutoff $E_c$ effectively controls the height of the energy deposition, i.e. how deep the lowest-energy electrons (which carry most of the energy) can penetrate the atmosphere. 
A beam with a harder spectral index (lower $\delta$ value) will spread the energy deposition deeper in the atmosphere than a softer (higher $\delta$ value). 

The pre-flare atmosphere that we use for all runs, as shown in many of our figures, {is a radiative equilibrium model \citep{AbbettHawley:1999,AllredKowalskiCarlsson:2015}}, with a {maximum temperature of 1~MK at the top of the loop, which has a total length of 10~Mm. The pre-flare transition-region pressure is about 0.22~dyne cm$^{-2}$.} Its details {may} have only minor relevance to the flare effects. For comparison the standard VAL-C model \citep{VernazzaAvrettLoeser:1981} has a transition-region pressure of about 0.2~dyne cm$^{-2}$.
This short time scale that we have adopted allows us to follow the impulsive development of ionization and mid-IR emission in the constant-density limit \citep{ShmelevaSyrovatskii:1973}, in which the flows resulting from the energy input cannot play a major role for lack of time in which to develop.
For each model, we used the output atmosphere at time intervals $\Delta t=0.1$s 
 to calculate the opacities $\kappa_\nu(H)$ and $\kappa_\nu(H^-)$, the contribution function CF, and brightness temperature $T_b$ (Eq. \ref{eq:kff} to \ref{eq:tb}) for wavelengths in the infrared: $\lambda=2$, 10, 50, 200 $\mu$m ($\nu=150$, 30, 6 and 1.5 THz, respectively).
%
\begin{figure*}
\resizebox{\hsize}{!}{\includegraphics[angle=0]{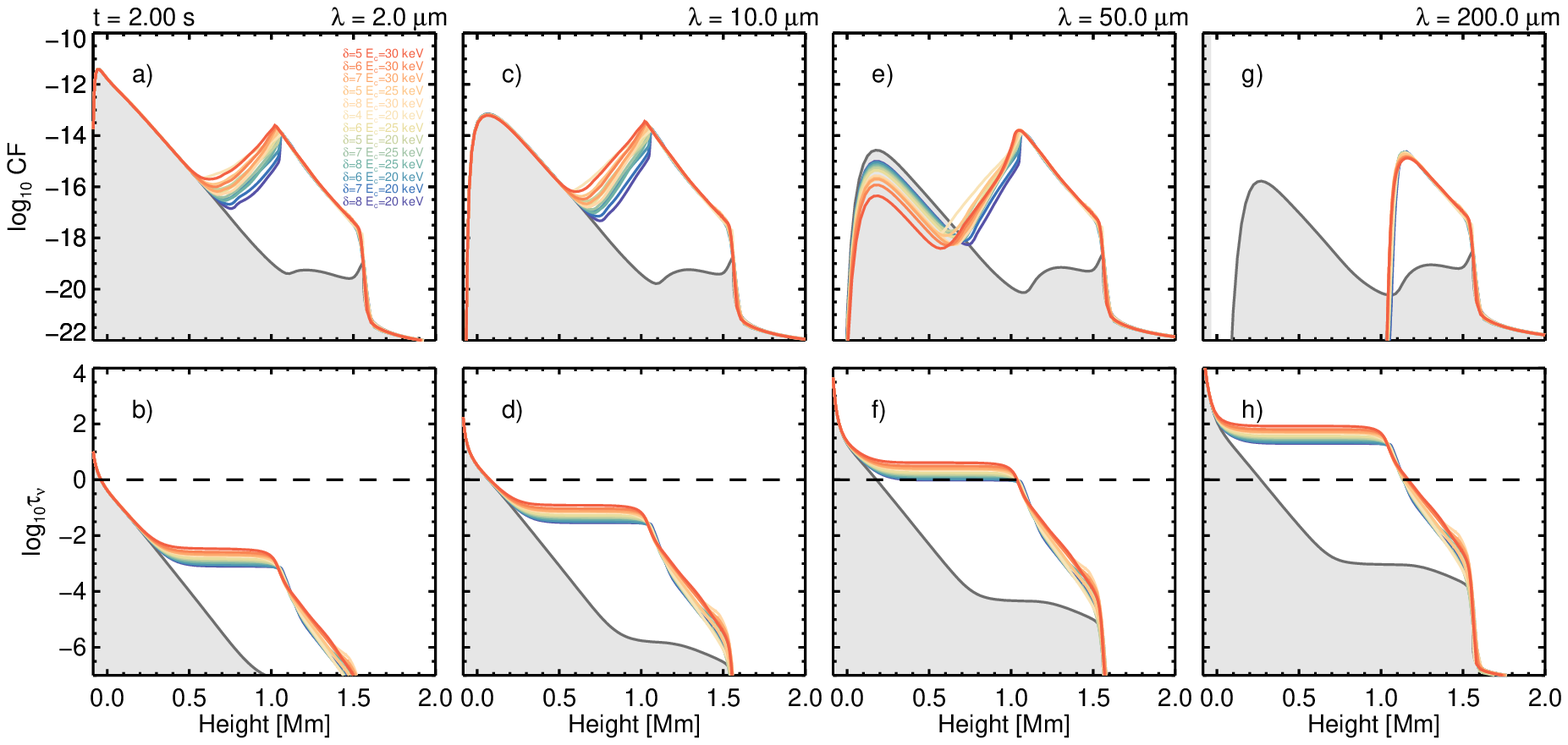}}
\caption{Atmospheric snapshots at peak energy deposition (CF and optical depth) for 2, 10, 50 and 200 $\mu$m.
The dashed line shows optical depth unity, which falls in the chromosphere at wavelengths longer than about 50~$\mu$m.
}
\label{fig:cfmulti}
\end{figure*}
\section{Results}

\subsection{Formation height and optical depth} \label{sec:formation}

In Figure \ref{fig:cf10um} we show the flaring atmospheric parameters from RADYN simulations and IR calculations at $\lambda=10\ \mu$m for all the models, at the time of maximum energy input ($t=2.0$ s). 
Different beam models deposit the energy in slightly different ways (see panel Fig. \ref{fig:cf10um}e), causing different atmospheric responses e.g. temperature and pressure (especially above 1 Mm). The chromospheric temperatures (Fig. \ref{fig:cf10um}a) rise from a few $10^3$ K to a few $10^4$ -- $10^5$ K between $0.7 < h < 1.5$ Mm, along with a large increase of the electron density (Fig. \ref{fig:cf10um}c), from about $10^{10}$ cm$^{-3}$ to a few $10^{13}$ cm$^{-3}$, as hydrogen becomes fully ionised (Fig. \ref{fig:cf10um}f), at and above the locations where the peak heating occurs. {There are no significant effects below $h~\approx~0.5$~Mm. The short heating timescales in these models is not sufficient to increase the temperature in the mid-chromosphere enough to result in strong backwarming by UV radiation and subsequent photospheric temperature enhancement \citep[e.g.][]{MachadoEmslieBrown:1978}.}

Qualitatively, the production of mid-IR (10~$\mu$m) emission is very similar for all of our models. During the simulated flares, the IR emission has an optically thin contribution from the flaring chromosphere superposed on an optically thick contribution from the undisturbed photosphere (Section~\ref{sec:formation}), with the flare IR emission increase due almost entirely to changes in the optically thin contribution.
The contribution function CF (Figure~\ref{fig:cf10um}b) shows an enhancement in the chromosphere, with a maximum just above 1~Mm, varying within 0.1~Mm depending on the beam model. 
The optical depth (Figure~\ref{fig:cf10um}d) increases by many orders of magnitude, but remains optically thin, reaching $0.1 < \tau_\nu < 0.01$, depending on the beam model. 
The enhancement of $\tau_\nu$, and consequently the CF is a direct result of the increase of the electron density rather than an increase in local temperature.

The main differences in CF and $\tau_\nu$ for the various beam models appear roughly at heights $0.6 < h < 1$~Mm. These differences can be traced to the effect of the beam on the hydrogen ionisation and consequent increase of the electron density. Beam models with higher low-energy cutoff $E_c$ values and harder spectral indices $\delta$ will have more electrons with higher energies, that can penetrate and deposit energy deeper in the chromosphere \citep[e.g.][]{Emslie:1978}, where the higher density of hydrogen provide a larger supply of electrons.

The beam model that gives the largest CF and $\tau_\nu$ has $E_c=30$ keV and $\delta=5$, which is able to heat deeper in the atmosphere, ionising more hydrogen than the other beam models and thus producing more free electrons. 
The weakest cases are $E_c=20$ keV, with both $\delta=7$ and $\delta=8$, meaning that, proportionally, most of the electrons collisionally stop higher in the chromosphere (about 0.1 Mm higher than the latter case), and hence produce fewer free electrons. 

We now characterize the IR emission at wavelengths $\lambda=2$, 50 and 200 $\mu$m, comparing with the results at 10~$\mu$m (Figure~\ref{fig:cf10um}). 
Figure~\ref{fig:cfmulti} shows the contribution function CF and the optical depth $\tau_\nu$ for all the beam models, at the four selected wavelengths. 
The CF at 2~$\mu$m and 10~$\mu$m (Figure \ref{fig:cfmulti}a and Figure \ref{fig:cfmulti}c) are very similar for all the models, and $\tau_\nu$ at 2 $\mu$m (Figure \ref{fig:cfmulti}b) is smaller than at 10~$\mu$m (Figure \ref{fig:cfmulti}d). 
At 50~$\mu$m the opacity regime in the chromosphere starts to change, with the optical depth reaching $\tau_\nu > 1$ (Figure \ref{fig:cfmulti}f). 
This affects the CF (Figure \ref{fig:cfmulti}e): {as the optical depth increases in the chromosphere past $\tau_\nu > 1$, the contribution to CF from the {\em undisturbed} photosphere is strongly reduced.} 
{As the hydrogen recombines following the cessation of flare energy deposition, the opacity regime reverts to being optically thin (this happens around $t = 6$~s in our simulations).}
At 50 $\mu$m, during the energy deposition phase, the flare emission is predominantly optically thick from the chromosphere.

At 200 $\mu$m, the chromospheric emission becomes very optically thick ($\tau_\nu > 10$) below $\approx 1$ Mm at $t=2.0$ s. In fact, as soon as H starts to ionise (around $t=1.6$ s), the electron density reaches $n_e \approx 10^{12}$ cm$^{-3}$, which is sufficient to make $\tau_\nu>1$. After the energy input, the chromosphere remains highly ionised, supplying enough free electrons to maintain $\tau_\nu>1$ until the end of the simulated flares, at $t=60$ s.
These differences in CF and $\tau_\nu$ for a range of IR wavelengths are a direct consequence of the ion free-free opacity, Eq. \ref{eq:kff}, which increases for lower frequencies (longer wavelengths). 

In our simulations, the change in the opacity regime, from optically thin to thick, occurs around 50 $\mu$m. However, the wavelength at which this change happens will depend slightly on the initial atmosphere and strongly on the intensity of the energy deposition. 
The density profile of the atmosphere defines the column depth for penetration of the electron beam, and hence the heating.
It also dictates the amount of hydrogen that can be ionised, and therefore, sets a limit to the largest electron density available.

\subsection{Comparison with semi-empirical atmospheric models}\label{sec:semi}
\begin{figure}
\resizebox{\hsize}{!}{\includegraphics[angle=0]{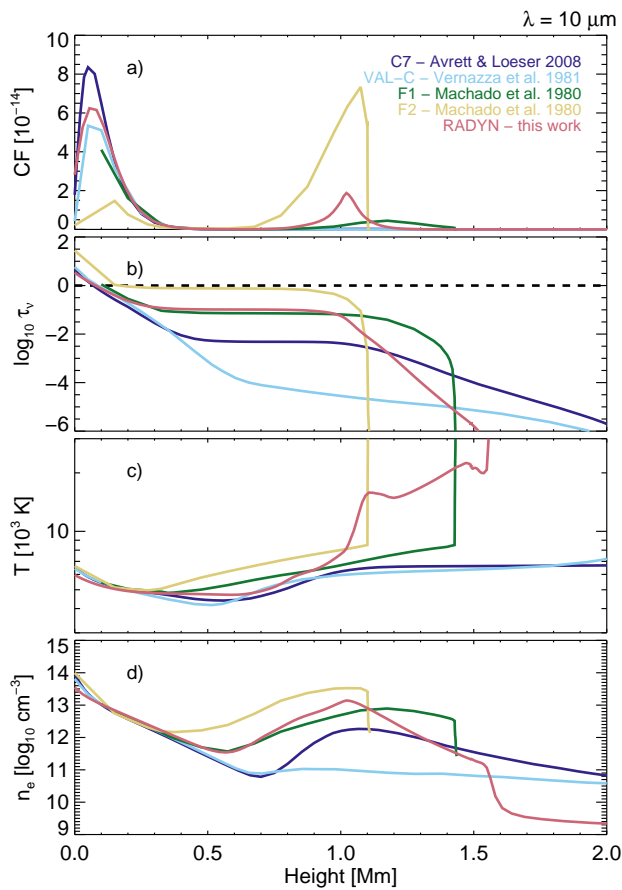}}
\caption{{Comparison between of our time-averaged RADYN calculations and semi-empirical atmospheric models (see details in Sect. \ref{sec:semi}) at 10 $\mu$m. a) Contribution function CF, b) optical depth $\tau_\nu$ (dashed line indicates $\tau_\nu=1$), c) Temperature $T$ and d) electron density $n_e$.}}
\label{fig:semi}
\end{figure}

{
Semi-empirical atmospheric models are often used to assess the formation height of the radiation under investigation or properties of the atmosphere \citep[e.g.][]{XuCaoLiu:2004,TrottetRaulinMacKinnon:2015}. Here we compare our calculations of the
time-averaged ($1<t<3$~s) contribution function CF and $\tau_\nu$ at 10~$\mu$m from our model with $\delta=5$ and $E_c=30$~keV, with the same properties obtained from the semi-empirical models for the quiet-Sun VAL-C \citep{VernazzaAvrettLoeser:1981} and C7 \citep{AvrettLoeser:2008}, and flare models F1 and F2 \citep{MachadoAvrettVernazza:1980}. CF and $\tau_\nu$ are presented in Figure \ref{fig:semi}, along with the temperature $T$ and electron density $n_e$ for each model. The reason for using time-averaged quantities from our calculations is simply to approach the time- and space-average nature of the semi-empirical models. Qualitatively, all models are in good agreement: all models present an optically thick ($\tau_\nu > 1$) contribution formed in the photosphere ($h < 0.1$~Mm), with an optically thin ($\tau_\nu \ll 1$) chromospheric contribution ($h \approx 1$~Mm) in the flare models. The location and intensity of the chromospheric CF of a given model is a direct consequence of its electron density profile.

The F2 model has a stronger CF in the chromosphere than in the photosphere, as is the case for F1 and our calculations. This effect comes directly from the model's larger chromospheric $n_e$ values, which in turn, comes from the larger hydrogen density between 0.5--1 Mm \citep{MachadoAvrettVernazza:1980}. The larger $n_e$ produces a larger optical depth, that approaches $\tau_\nu \approx 1$ in the chromosphere, essentially blocking the photospheric contribution. Note that this does not occur for the `weaker' F1 model, and its photospheric contribution is in very good agreement with the quiet-Sun models VAL-C and C7, and also with our RADYN calculations.

Note the excellent agreement of our CF and $\tau_\nu$ calculations for the models C7 and F2 with the calculations by \cite{TrottetRaulinMacKinnon:2015}, who used the PAKAL radiative transfer code \citep{De-la-LuzLaraMendoza-Torres:2010,De-la-LuzLaraRaulin:2011}. We have also calculated FLA and FLB flare models from \cite{MauasMachadoAvrett:1990}, and note that they do not produce a strong enhancement of the IR emission against the quiescent photospheric emission, given that both models have a lower $n_e$ in the chromosphere compared to F1 and F2. 
}
\subsection{Sources of IR opacity enhancement}
The near-- and mid--IR emission in the quiet Sun is dominated by the H$^-$ opacity near disk center. The physical height of the $\tau_\nu = 1$ layer falls just longward of the H$^-$ absorption edge, slightly below the height of the visible photosphere (the ``opacity minimum''), but increases monotonically for longer wavelengths.
It reaches the height the visible photosphere again roughly at the temperature minimum region, near 200~$\mu$m. 

\begin{figure*}
\resizebox{\hsize}{!}{\includegraphics[angle=0]{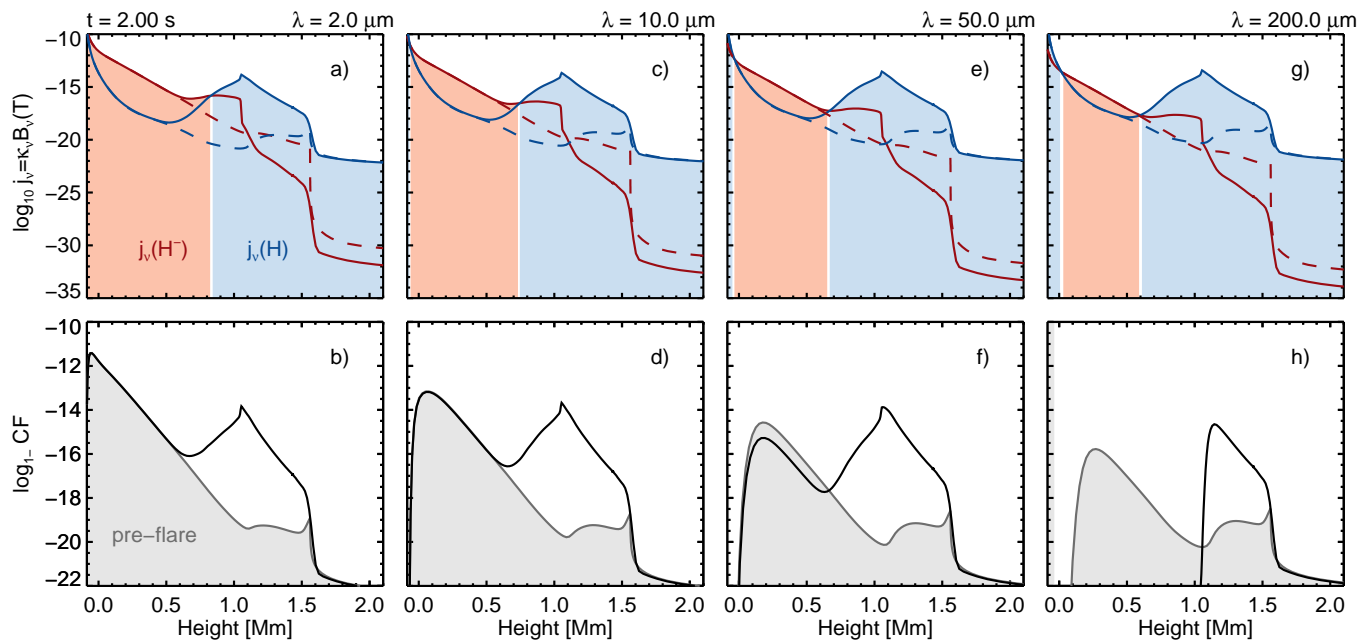}}
\caption{Upper row: neutral free-free (H$^-$) and ion free-free emissivities for the model with $E_c=20$~keV and $\delta=5$, observed at $t=2.0$~s, at each of four wavelengths.
Red shading indicates neutral free-free dominance, and blue shading ion free-free.
Lower, the corresponding contribution functions, showing the {enhancement of the} optical depth at longer wavelengths as the ion free-free opacity increases.
For these parameters, panel (f) shows the chromosphere to become completely opaque by 200~$\mu$m. {The dashed lines show the pre-flare emissivities.}
}
\label{fig:kappas}
\end{figure*}

Under flaring conditions, the flare energy deposition ionises hydrogen, helium, and the metals, which supplies free electrons that enhance the ion free-free opacity and creates the IR flare excess emission. 
{The interplay of these two sources of opacity as a function of wavelength shown in Figure~\ref{fig:kappas} evidence the dominance of ion free-free opacity (blue) in the chromosphere over the contribution from the undisturbed photospheric H$^-$ opacity (red).}

Note that RADYN explicitly calculates the radiative transfer necessary to estimate the contributions from hydrogen and helium, including continua and many lines \citep[e.g.][]{AllredKowalskiCarlsson:2015}. Helium does not play an important role as a source of electrons, but its ionization to He~{\sc ii} and then He~{\sc iii} contributes to defining the electron temperature. Thus, hydrogen ionization generally dominates the supply of free electrons.
\subsection{Synthetic IR flare spectrum}\label{sec:spec}
Qualitatively, we can describe the development of the flare emission as follows. 
Initially hydrogen ionizes at the depths the incident electrons can reach, with the height of full ionization moving downward as the energy input continues.
As this happens, the chromospheric ionization contributes an increasing opacity. Initially, throughout the mid-IR the chromosphere remains optically thin and the photosphere remains visible through the chromospheric contribution.
However, ultimately the electron column depth becomes sufficient to produce {a larger} optical depth, starting with the longest wavelengths. 

\begin{figure}
\resizebox{\hsize}{!}{\includegraphics[angle=0]{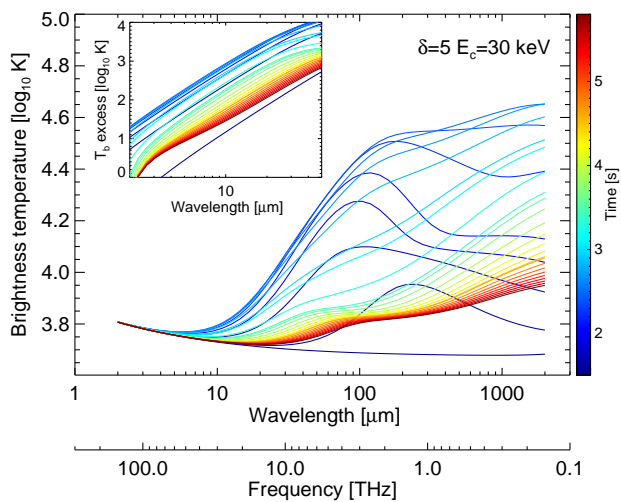}}
\caption{{Evolution of the IR flare $T_b$ spectrum for the model with $\delta=5$ and $E_c=30$ keV. For $\lambda > 50 \mu$m the brightness temperature tends to reflect the actual electron temperature in the optically-thick upper chromosphere.
The inset planel shows the flare excess emission, after subtracting the pre-flare spectrum. This highligths the positive slope of the short-wavelength branch, $T_b(\nu) \propto \nu^2$, that can be identified with the Rayleigh-Jeans law expected from an optically-thin slab.}
}\label{fig:irspec}
\end{figure}

\begin{figure}
\resizebox{0.9\hsize}{!}{\includegraphics[angle=0]{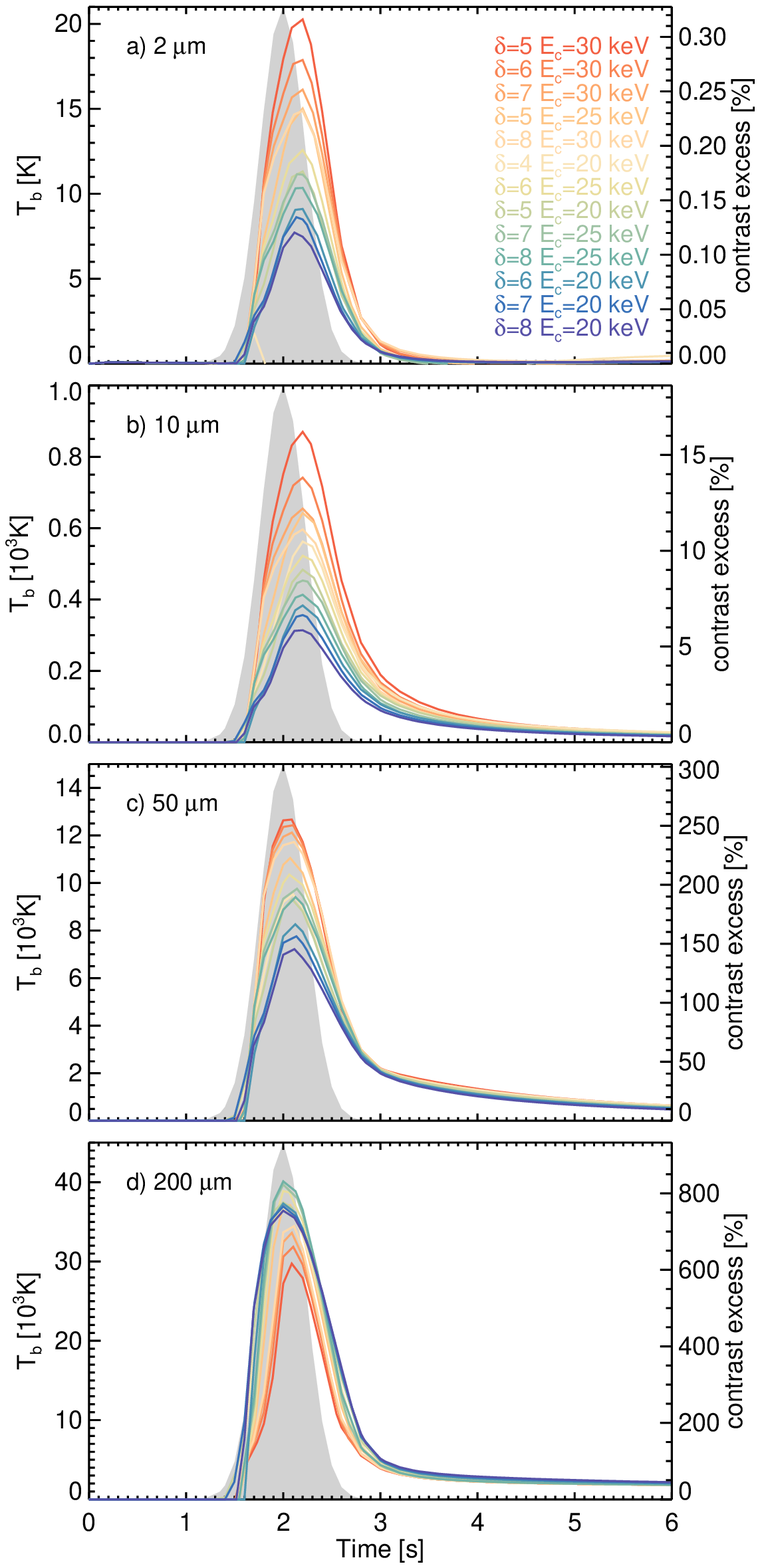}}
\caption{Evolution of the brightness temperature $T_b$ calculated at 2, 10, 50, 200 $\mu$m for all 13 models. The gray area schematically shows the time profile of energy input {F11t, identical for all models (see Sect. \ref{sect:models}).}
}
\label{fig:lc}
\end{figure}

{Figure \ref{fig:irspec} shows the evolution of the IR flare $T_b$ spectrum for the model with $\delta=5$ and $E_c=30$ keV. The shape of the $T_b$ spectrum deviates from the typical form for an isothermal source \citep{Dulk:1985}. For $\lambda > 50 \mu$m the brightness temperature tends to reflect the actual electron temperature in the non-uniform, optically-thick upper chromosphere. Note that the spectrum remains enhanced after the energy input has finished around $t = 3$~s, especially at longer wavelengths. The short-wavelength branch can be identified with the Rayleigh-Jeans law expected from an optically-thin slab, $T_b(\nu) \propto \nu^2$, more easily seen in the inset panel of Figure \ref{fig:irspec}, which shows the flare emission after subtracting the pre-flare spectrum.
}
Figure \ref{fig:lc} shows the evolution of the brightness temperature $T_b$ and contrast excess CE~$=\Delta I/I_\lambda(0)$ for 2, 10, 50, and 200 $\mu$m. For all models, $T_b$ shows an impulsive rise, quickly decreasing after the energy deposition finishes. 
Neither $T_b$ nor CE are very significant at 2 $\mu$m (CE $< 1$\%), but they increases substantially for longer wavelengths and effectively increase the observability of the process.
At the shorter wavelengths (2 and 10 $\mu$m in Figure \ref{fig:lc}), where the chromospheric emission is optically thin, there is a small time difference between the peak of the energy deposition and the peak of the emission. 
This is because the ionisation timescale is shorter than the recombination timescale, allowing the accumulation of free electrons in the chromosphere. 
As the energy deposition diminishes, the ionisation rate decreases and the recombination rate starts to dominate, so that the IR emission fades. 

\begin{figure}
\resizebox{\hsize}{!}{\includegraphics[angle=0]{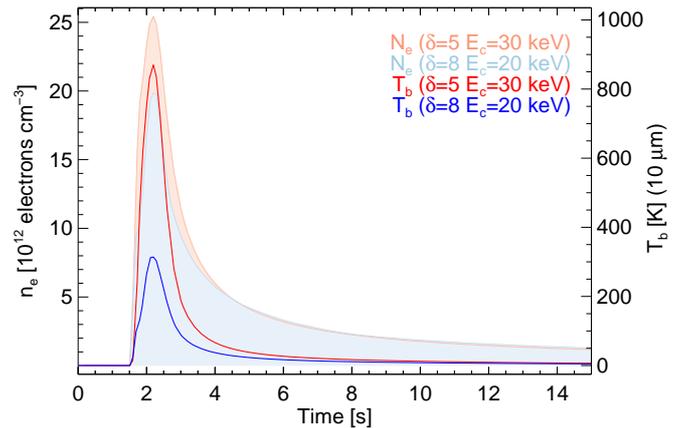}}
\caption{Time evolution of the electron density $n_e$ in the chromosphere at the height where CF(10$\mu$m) is maximum, for the models $\delta=5$ and $E_c=30$ keV (pale red) and $\delta=8$ and $E_c=20$ keV (pale blue). The brightness temperature $T_b$ at 10~$\mu$m {for each case is shown in dark red and dark blue curves, and indicates} that mid-IR emission is a proxy for the $n_e$ and H ionisation in the chromosphere.}
\label{fig:nelevol}
\end{figure}

{The IR emission $T_b$ directly depends on the electron density $n_e$ in the chromosphere. Figure \ref{fig:nelevol} shows the $T_b(10\mu$m$)$ from the models that produced the maximum and minimum $T_b(10\mu$m$)$ values, respectively $\delta=5$ and $E_c=30$~keV (red curve) and $\delta=8$ and $E_c=20$~keV (blue curve), compared with the electron density $n_e$ 
{at the height corresponding to the peak of CF(10~$\mu$m)}. Although the thickness of the ionised chromosphere and its temperature structure evolve with time, the short-wavelength IR emission is a clear proxy for the chromospheric electron density, and thus ionisation state of hydrogen. The direct implication of this finding is that mid-IR observations can be used to track the region where energy is being deposited, without the extra complications of formation and optical depth of spectral lines or recombination continua, often associated with typical flare observations, e.~g. H$\alpha$ \citep{KuridzeMathioudakisSimoes:2015}, Mg {\sc ii} \citep{KerrSimoesQiu:2015}, or white-light continuum \citep{KowalskiAllredDaw:2016}.}

At longer wavelengths (50~and 200~$\mu$m in Figure \ref{fig:lc}), the emission is optically thick, and $T_b$ saturates at the local electron temperature $T$. As the heating increases, hydrogen ionises higher in the chromosphere, shifting the $\tau_\nu=1$ layer upwards, where the electron temperature is progressively higher, hence $T_b$ increases. 
When the energy deposition finishes, hydrogen recombines, the $\tau_\nu=1$ layer moves back to lower altitudes, and $T_b$ decreases. Note that at these longer wavelengths, the decay time scale following the end of energy deposition is longer than for the shorter wavelengths.

The impulsive IR emission is  directly linked to the evolution of the H ionisation which, in our simulations, is determined by the heating function used. \citet{PennKruckerHudson:2016} observations suggested impulsive heating timescales shorter than $\sim$4~s, with an exponential cooling with a $e$-folding time of 10~s. It is likely that the IR sources are not resolved in the observations, and that multi-thread sources on fine scales must  be considered \citep{HoriYokoyamaKosugi:1998,ReepWarrenCrump:2016}. {The presence of multiple, unresolved, fine-scale events might help to explain the extended
temporal emission that is observed by \cite{PennKruckerHudson:2016}.}

\section{Comparison with observations}~\label{sect:data}
	
\begin{figure*}
\resizebox{0.8\hsize}{!}{\includegraphics[angle=0]{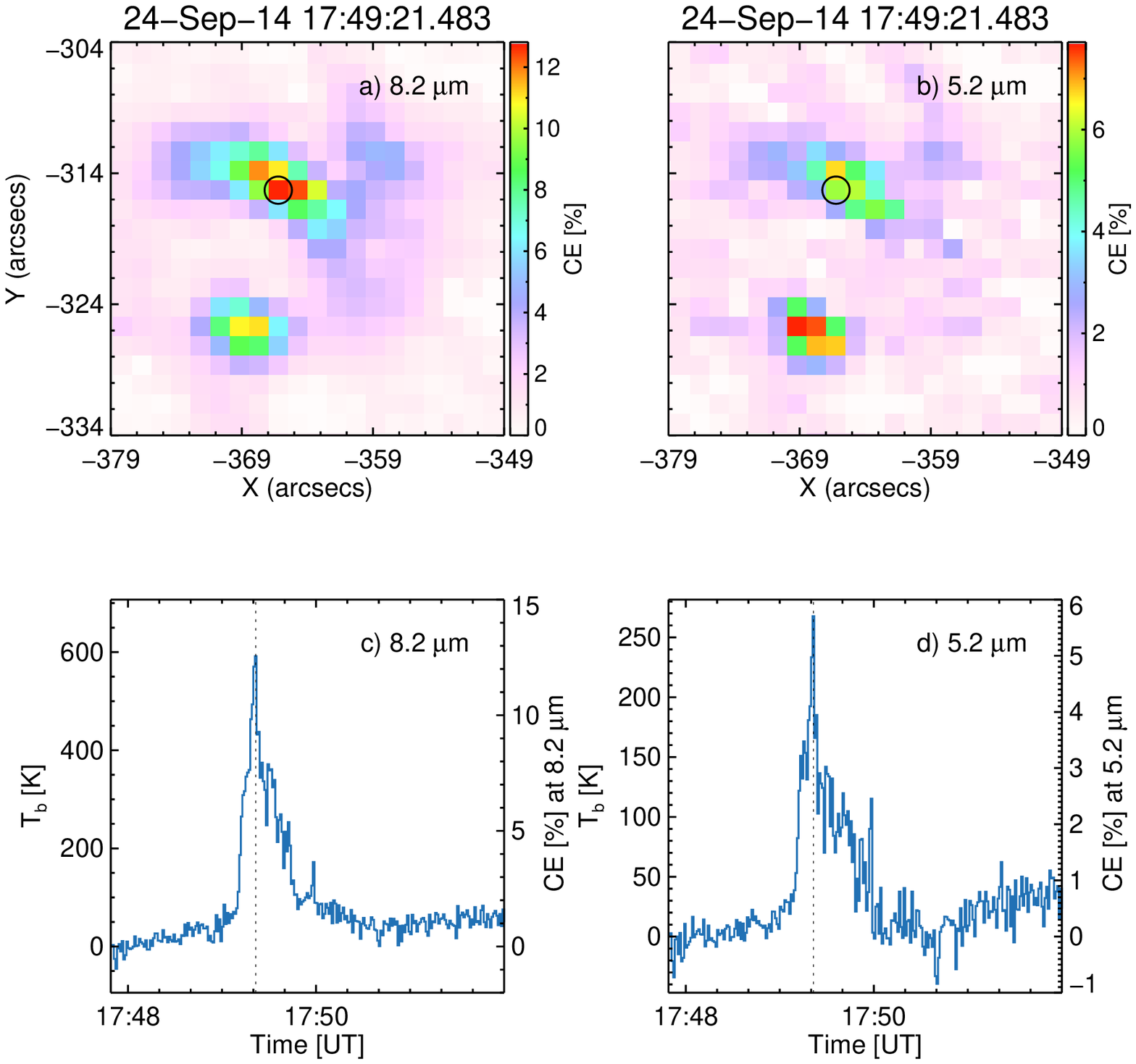}}
\caption{Data from \citet{PennKruckerHudson:2016}. The upper panels show preflare-subtracted flare images in a  one-second snapshot at the indicated time.
The lower panels show light curves from the region identified with the black circle.}
\label{fig:data}
\end{figure*}

In a recent observation \citet{PennKruckerHudson:2016} obtained the first two-frequency spectrophotometry in the mid-IR, in broad wavelength bands centered at about 5.2 and 8.2~$\mu$m.
We will consider here how well our impulsive modelling provides a qualitative match to these observations. Figure~\ref{fig:data} summarizes these observations, obtained at the NSF's McMath-Pierce (McMP) Solar Facility telescope with quantum well infrared photodetector (QWIP) imaging arrays, capable of sub-second time resolution.
These unique observations have clearly defined properties, i.e. double ``footpoint'' source, close to the white-light continuum and hard X-ray sources, and rapid variability.
 Fig. \ref{fig:data} shows maps of the contrast excess at the peak of IR emission (17:49:21~UT) during the event SOL2014-09-24 (C7.0; S10E31). 
These contrast maps were obtained by subtracting a time average of 6 images before the flare from the image at the time of interest. 
Two flare sources are clearly visible with highly significant $T_b$ and contrast excesses CE of a few percent, greater at the longer wavelength.
Light curves of a single pixel in the northern source are also shown in Fig. \ref{fig:data}, along with brightness temperature $T_b$ values for the flare excess for both wavelengths. 

At 8.2 $\mu$m, at the time of the maximum emission the excess flare brightness $T_b $ is observed to be $\approx 600$ K, for a contrast of $\approx 13$\%. 
At 5.2 $\mu$m the signal is weaker relative to the background, at $T_b\approx 270$ K, and contrast $\approx 5.7$\%. 
These imaging observations have well-defined absolute calibration from the preflare observations of the quiet Sun \citep[e.g.][]{1973asqu.book.....A}.
As noted by \citet{PennKruckerHudson:2016}, these values roughly match the expectation from a Rayleigh-Jeans spectrum, consistent with a warm, optically thin, chromospheric slab source.

\begin{figure}
\resizebox{\hsize}{!}{\includegraphics[angle=0]{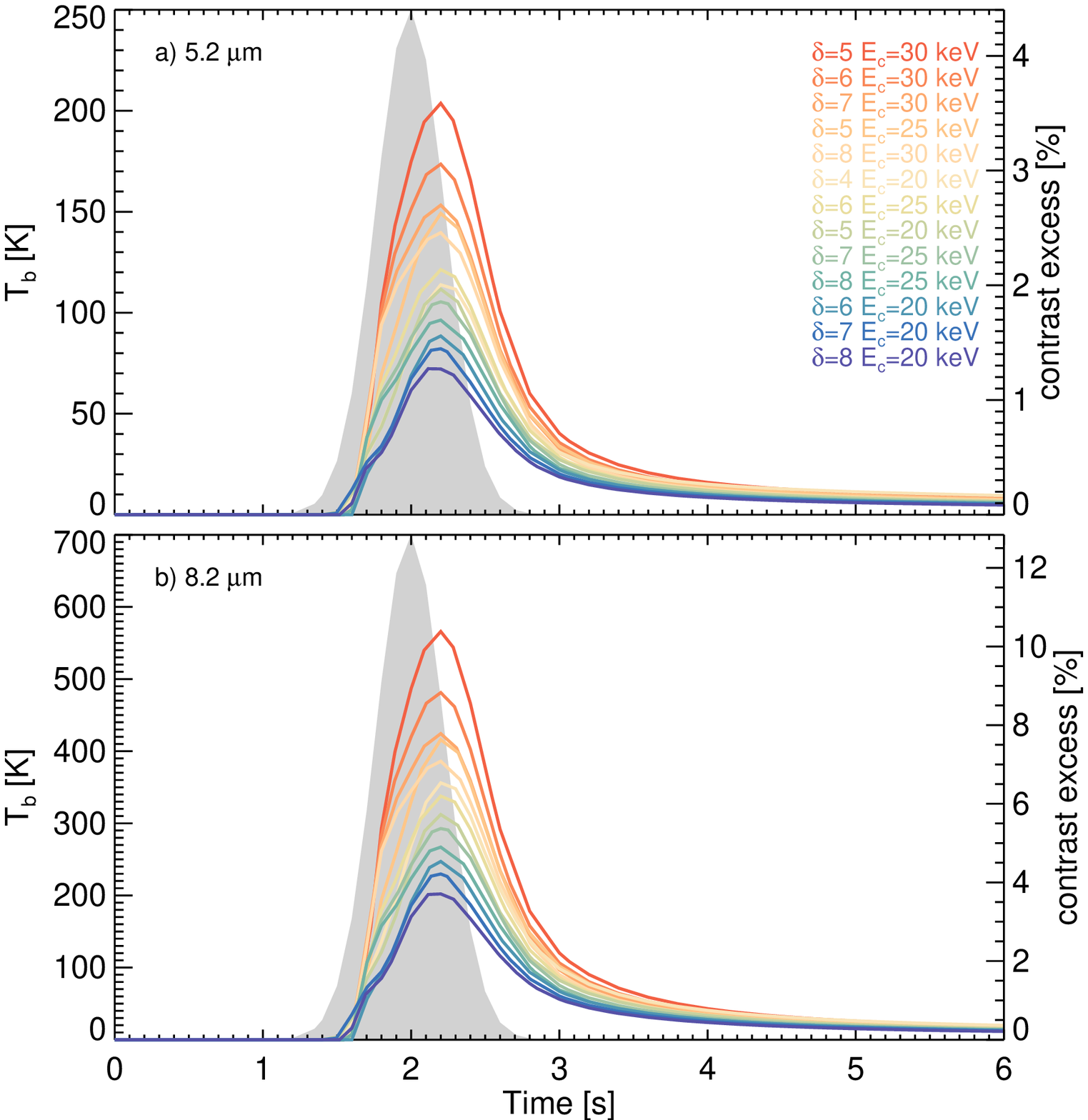}}
\caption{Same as Figure \ref{fig:lc}, but at wavelengths (a) 5.2 and (b) 8.2 $\mu$m for comparison with \citet{PennKruckerHudson:2016} observations. The gray area schematically shows the time profile of energy input {F11t, identical for all models (see Sect. \ref{sect:models}).}}
\label{fig:lcqwip}
\end{figure}

RHESSI detected hard X-rays, and HMI observed white-light emission from this event.
From the hard X-ray observations and assuming a standard thick-target for the spectrum fitting, \citet{PennKruckerHudson:2016} report a spectral index of $\delta=3.8$, and estimate an energy deposition rate of $1.1 \times 10^{11}$ erg s$^{-1}$ cm$^{-2}$, for electrons above $E_c=50$ keV. These values motivated our choices for the RADYN models presented in this paper, and calculations of $T_b$ and contrast for the \citet{PennKruckerHudson:2016} wavebands 5.2 and 8.2 $\mu$m are shown in Figure~\ref{fig:lcqwip}. 
These capture the main features of the observation, specifically the contrasts, the spectral ratio, and the time scales {(considering that the IR sources are not spatially resolved in the observations, see Sect.~\ref{sec:spec})}. In particular the Rayleigh-Jeans-like spectral ratio appears as a natural consequence of the small chromospheric optical depth at these wavelengths. 


This straightforward comparison indicates that our simulations capture the main mechanisms involved in the production of the IR emission in solar flares. 
This strongly suggests that the IR emission is {due to thermal ion free-free} emission generated by the ion free-free mechanism, as proposed by \citet{OhkiHudson:1975}, \cite{KasparovaHeinzelKarlicky:2009}, and \citet{TrottetRaulinMacKinnon:2015}, and tentatively confirmed by the \citet{PennKruckerHudson:2016} observations. 
The necessary content of free electrons are mainly supplied by the ionisation of hydrogen in the chromosphere. For these wavelengths, the observations are consistent with optically thin ion free-free emission, as described by  our simulations.

These consistencies are heartening and confirm our interest in the constant-density limit, and in the parameter space we have sampled.
Based upon these models, we can anticipate much more diagnostic power from mid-IR to submm ranges (especially for sources nearer the limb, where greater heights in the chromosphere can contribute).
Although the duration of our simulated flares is shorter than the overall duration of the observed event, the general properties are also in excellent agreement: a fast, impulsive rise of the mid-IR emission and also a quick, but slightly slower, decrease. 
The second, slower decrease of the emission we have discovered in these models, due most probably to hydrogen ionisation left over in the upper chromosphere, could not readily be isolated in the Penn et al. observations.

\section{Discussion}~\label{sect:discussion}

Our modeling has assumed that the energy transport is entirely by electron beams, which may only be part of the story \citep[e.g.][]{FletcherHudson:2008}. 
{It is also possible to heat the chromosphere during flares via the dissipation of Alfvenic waves, mainly via ion-neutral friction \citep{EmslieSturrock:1982,ReepRussell:2016,KerrFletcherRussell:2016}.}
HXR observations demonstrate that there are fast electrons present in the chromosphere though whether in a the form of a beam from the corona or not is unknown, {and of course, both mechanisms may contribute}. Therefore the presence of free electrons during solar flares to enhance the IR opacity in the chromosphere is not particularly model-dependent.

{The non-thermal electrons play an important role in enhancing the excitation and ionisation of H in the chromosphere. In general, non-thermal collisional rates dominate the H excitation and ionisation from level $n=1$ \citep[see e.g.][]{KarlickyKasparovaHeinzel:2004,KasparovaVaradyHeinzel:2009}, but the H ionisation from level $n=2$ and above are still controlled by thermal rates. Since the presence of non-thermal electrons enhances H ionisation and excitation in the chromosphere, and the IR flare emission is essentially dependent on high electron density values ($n_e > 1\times 10^{12}$~cm$^{-3}$), one would expect HXR signatures from non-thermal electrons and the IR emission to be very well associated, as it was in fact observed by \cite{PennKruckerHudson:2016}.}
 
The height distribution of the energy deposition does matter, in that a large IR enhancement requires increased ionization in a dense atmospheric layer but the short mid-IR wavelengths (2-10$\mu$m) being optically thin, track the chromospheric `total electron content' and do not provide information on the chromospheric electron density distribution.

The departures from the Rayleigh-Jeans law seen in Figure~\ref{fig:irspec} suggest that future flare observations with the Atacama Large Millimeter/sub-millimeter Array \citep[ALMA,][]{2016arXiv160100587W} will help to {find the temperature structure of the flaring chromosphere, providing valuable information to identify the energy transport mechanisms. Note, however, that at mm-waves the contribution from synchrotron emission from non-thermal electrons may become important or even dominant \citep[e.g.][]{TrottetRaulinKaufmann:2002,LuthiMagunMiller:2004,LuthiLudiMagun:2004,TrottetRaulinGimenez-de-Castro:2011,Gimenez-de-CastroCristianiSimoes:2013,TsapSmirnovaMorgachev:2016,2017SoPh..292...21F}.}

We have shown that the mid-IR (2--10~$\mu$m) continuum intensity can vary on time scales close to those of the energy input, in agreement with the findings of \citet{KasparovaHeinzelKarlicky:2009} {for the sub-mm/mm range}. However, it persists on a longer time scale after the end of the energy input, and we attribute this to the accumulation of free electrons in the chromosphere {due to the non-equilibrium ionization.}. 
The mid-IR continuum intensity is then a good proxy for the evolution of (primarily) hydrogen ionization, at around 1~Mm in these electron-beam-driven models, and also for the chromospheric electron content.

\section{Conclusions}~\label{sect:conclusions}

Using radiative hydrodynamic simulations, we have confirmed that the enhanced mid-infrared continuum in solar flares forms mainly in the chromosphere, {via} the ion free-free mechanism. 
We confirm with these models the rapid ionization and recombination of hydrogen needed to allow the mid-IR to exhibit time scales close to those of an impulsive energy input (the Shmeleva-Syrovatskii ``constant density'' approximation).
The rapid ionisation of neutral hydrogen due to the energy input supplies the free electrons for the enhancement. Ionisation of He {\sc i} to He {\sc ii} can also contribute free electrons in the upper chromosphere.  The brightness temperature evolution of the mid-IR continuum emission is a good proxy for the chromospheric `total electron content', but its peak slightly lags the peak of the energy input. At higher altitudes in the chromosphere, seen in the simulations in the continuum longward of 200~$\mu$, we see a substantially different decay profile.

The IR continuum emission is optically thin from the shorter wavelengths (2 $\mu$m) to about 50 $\mu$m, formed in the chromosphere. The $\tau_\nu=1$ layer throughout the mid-IR remains in the undisturbed photosphere for our parameter choices. More detailed spectral observations in this (and longer) wavelengths could define the development of electron content with better detail, possibly sufficient to clarify the energy-transport mechanism. 
In the meantime a simple uniform slab model can be used to assess the electron density where the IR continuum is formed: $T_b=\tau_\nu T=\kappa_\nu^{ff}LT \propto n_e^2L T^{-1/2}$, for a slab with thickness $L$.

On the other hand, for $\lambda > 50 \mu$m, the emission becomes optically thick for our parameter ranges.
The $\tau_\nu=1$ layer moves to the upper chromosphere, and observations should give a direct measurement of the kinetic temperature of the electrons, since $T_b=T$ for $\tau_\nu >> 1$.
This temperature is highly dependent upon model parameters, specifically for the treatment of energy transport.

Observationally, we expect great progress from continuum observations via continuing mid-IR observations, from  observations in the near-IR with new large-aperture telescopes such as the 4-m DKIST, and from sub-mm/mm observations with ALMA.

\begin{acknowledgements}

The research leading to these results has received funding from the European Community's Seventh Framework Programme (FP7/2007-2013) under grant agreement no. 606862 (F-CHROMA). PJAS and LF acknowledge support from grant ST/L000741/1 made by the UK's Science and Technology Facilities Council. GSK acknowledges the financial support of a PhD research scholarship from the College of Science and Engineering at the University of Glasgow. HSH acknowledges support from NASA under contract NAS 5-98033 for RHESSI. GGC acknowledges CNPq (contract 300849/2013-3). We thank Dr. M. Carlsson for developing, and sharing with us, his RADYN code. The NSO is operated by AURA, Inc. under contract to the National Science Foundation (NSF). We acknowledge the use of colour-blind safe and print-friendly colour tables by Paul Tol (\verb+http://www.sron.nl/~pault/+). We also would like to thank the reviewer P.~Heinzel for helpful comments and corrections.

\end{acknowledgements}


\end{document}